\documentclass[conference,10pt]{IEEEtran}
\ifCLASSINFOpdf
\else
\fi

\def\P{{\mathcal{P}}}
\def\E{{\mathbb{E}}}
\def\pe{{\epsilon}}
\def\L{{V}}
\def\l{{v}}
\def\D{{P}}

\usepackage{amsmath,amssymb,amscd,amsthm}
\usepackage[english,ruled,vlined,linesnumbered]{algorithm2e}
\usepackage{array}
\usepackage[utf8]{inputenc}

\usepackage{etex}
\usepackage{graphicx}
\usepackage{tikz,pgfplots}
\usepackage{float}

\usepackage{rotating}
\usepackage{multirow}
\usepackage{tabularx}

\usepackage{tocbasic}
\usepackage{ifthen}
\usepackage{subscript}
\usepackage{verbatim}

\definecolor{light}{gray}{.75}

\newlength\figureheight
\newlength\figurewidth
\usetikzlibrary{arrows,chains,matrix,positioning,scopes,backgrounds,fit}
\makeatletter
\tikzset{join/.code=\tikzset{after node path={%
\ifx\tikzchainprevious\pgfutil@empty\else(\tikzchainprevious)%
edge[every join]#1(\tikzchaincurrent)\fi}}}
\makeatother
\tikzset{>=stealth',every on chain/.append style={join},
         every join/.style={->}}
\tikzstyle{labeled}=[execute at begin node=$\scriptstyle,
   execute at end node=$]
 \pgfplotsset{compat=1.8
 every axis/.append style={
 line width=1.25pt,
 tick style={line width=1.25pt, color=black, line cap=round}
 }
}

\renewcommand{\vec}[1]{\ensuremath{\mathbf{#1}}}
\newcommand{\M}[1]{\ensuremath{\mathbf{#1}}}
\newcommand{\Set}[1]{\ensuremath{\mathcal{#1}}}

\newcommand{\rate}{\ensuremath{r}}

\newtheoremstyle{def}
  {2ex}
  {2ex}
  {\normalfont\itshape}
  {}
  {\normalfont\bfseries}
  {\newline}
  { }
  {\thmname{#1}\thmnumber{ #2} (\thmnote{#3})}

\newtheoremstyle{the}
  {2ex}
  {2ex}
  {\normalfont\itshape}
  {}
  {\normalfont\bfseries}
  {\newline}
  { }
  {\thmname{#1}\thmnumber{ #2}}

\newtheoremstyle{col}
  {2ex}
  {2ex}
  {\normalfont}
  {}
  {\normalfont\bfseries}
  {\newline}
  { }
  {\thmname{#1}\thmnumber{ #2}}

\SetKwInput{KwPrep}{Initialize}

 \renewcommand\appendix{\par 
   \setcounter{chapter}{0}
   \setcounter{section}{0}%
   \setcounter{subsection}{0}%
   \setcounter{figure}{0}%
   \renewcommand\thechapter{\Alph{chapter}}%
   \renewcommand\thefigure{\Alph{section}.\arabic{figure}}
 } 

\begin{document}
%
\title{Analyzing Finite-length Protograph-based\\ Spatially Coupled LDPC Codes\vspace{-2mm}}

\author{
\IEEEauthorblockN{Markus Stinner}
\IEEEauthorblockA{
	Institute for Communications Engineering\\
	Technische Universit\"at M\"unchen, Germany\\
	\texttt{markus.stinner@tum.de}
	}
\and
\IEEEauthorblockN{Pablo M. Olmos}
\IEEEauthorblockA{
	Departamento de Teor\'{i}a de la Se\~{n}al y Comunicaciones\\
	Universidad Carlos III de Madrid, Spain\\
	\texttt{olmos@tsc.uc3m.es}
	}
}

\maketitle

\begin{abstract}
Eligible for student paper award.
The peeling decoding for spatially coupled low-density parity-check (SC-LDPC) codes is analyzed for a binary erasure channel.
An analytical calculation of the mean evolution of degree-one check nodes of protograph-based SC-LDPC codes is given and an estimate for the covariance evolution of degree-one check nodes is proposed in the stable decoding phase where the decoding wave propagates along the chain of coupled codes.
Both results are verified numerically.
Protograph-based SC-LDPC codes turn out to have a more robust behavior than unstructured random SC-LDPC codes.
Using the analytically calculated parameters, the finite-length scaling laws for these constructions are given and verified by numerical simulations.

\end{abstract}


%
\IEEEpeerreviewmaketitle

\section{Introduction}

{\let\thefootnote\relax\footnotetext{
Markus Stinner was supported by an Alexander von Humboldt Professorship endowed by the
German Federal Ministry of Education and Research. Pablo M. Olmos  was supported  by Spanish government MEC TEC2012-38800-C03-01. The authors would like to thank Gerhard Kramer,  for the fruitful discussions.}} \par

\IEEEPARstart{S}patially coupled low-density parity-check (SC-LDPC) codes are known to achieve capacity over  binary-input memoryless symmetric (BMS) channels under belief propagation (BP) decoding \cite{KudekarBMS}.
SC-LDPC code ensembles are constructed by coupling $L$ $(l,r)$-regular LDPC codes, each one of length $M$ bits, together with appropriate boundary conditions.
When $M$ tends to infinity and $L$ is large, SC-LDPC codes exhibit a  BP threshold arbitrarily close to the maximum-a-posteriori (MAP) threshold of the $(l,r)$-regular ensemble \cite{KudekarBMS,Lentmaier10}.


Several constructions for SC-LDPC codes have been proposed  but, in many cases, these constructions have been chosen  to simplify the analysis of performance rather than to construct strong codes.
For uncoupled LDPC ensembles, it is well known that construction by means of protographs gives important practical advantages with respect to the random construction \cite{Urbanke08-2,Divsalar09}. 
However, protograph LDPC ensembles are a class of multi-edge type LDPC codes that are hard to analyze. 
Indeed, proofs for achieving capacity over BMS channels \cite{KudekarBMS} and finite-length performance analyses \cite{Olmos13, Olmos13-3} have been proposed only for random SC-LDPC codes. 

In this paper we show that finite-length protograph-based  codes provide better error rates than random constructions in both the waterfall and the error floor regions.
To this end, we extended the finite-length analysis recently proposed in \cite{Olmos13, Olmos13-3} for random  SC-LDPC ensembles to the protograph construction. 
Analysis of the finite-length performance of  LDPC codes in the waterfall region  is typically addressed over the binary erasure channel (BEC) where scaling laws relating the finite-length code performance and the LDPC code parameters can be analytically computed \cite{Urbanke09}.
For the BEC, we consider a formulation related to belief propagation called peeling decoding (PD) \cite{Urbanke08-2}.
PD iteratively removes variable nodes from the Tanner graph whose value is known, which yields a sequence of graphs whose statistics define both the asymptotic and  finite-length properties of the code \cite{Luby01}.
An estimate of the PD error probability is obtained based on the average evolution of the number of degree-one (deg1) check nodes in the graph and the variance around this average \cite{Urbanke09}.

In this paper, we show how to compute  the average evolution of deg1 check nodes for protograph-based SC-LDPC codes under PD.
In addition, for both the random and protograph constructions, we  propose an accurate method to  estimate the variance around the computed expected evolution.
Our analysis shows that the performance gain obtained by the protograph SC-LDPC construction can be explained by a robust expected graph evolution in which there exists a high fraction of deg1 check nodes at those instants where the decoding process is exposed to failures.
The protograph scaling law is obtained  with a simple parameter correction in the SC-LDPC scaling law proposed in \cite{Olmos13, Olmos13-3}.

\section{Constructing SC-LDPC Code Ensembles}\label{sec:SC-LDPCensembles}
We consider classes of  SC-LDPC code ensembles based on the $(l,r)$ regular LDPC ensembles.
In this section, we present the random construction proposed in \cite{KudekarBMS,Olmos13} and its counterpart based on protographs.
The random and protograph ensembles are denoted by $(l,r,L)$ and $(l,r,L)_{\P}$, respectively.

\vspace{-2mm}
\subsection{The $(l,r,L)$ ensemble}
Consider $L$ uncoupled $(l,r)$-regular LDPC codes of code length $M$, where $r$ is the check degree and $L$ is the variable degree, $r\leq l$.
Each code has $M$ variables and $\frac{l}{r}M$ check nodes.
The codes occupy $L$ consecutive positions.
We generate the SC-LDPC $(l,r,L)$ ensemble by spreading $l-1$ edges per variable node along consecutive positions.
Each variable node at position $u$ is connected to a check node at positions $u, u+1,\ldots,u+l-1$.
At each position, the check node is chosen  at random.
The code has $M$ variable nodes placed at positions $1,\ldots,L$ and there are $L+(l-1)$ positions with check nodes of non-zero degree.  The
design rate
tends to $1-l/r$, the uncoupled code rate, 
when $L\rightarrow\infty$. 

\subsection{The $(l,r,L)_{\P}$ ensemble}
Codes based on protographs were introduced by Thorpe in \cite{Thorpe03}.
Small Tanner graphs called protographs are used as templates for a large code construction: the small protograph is first copied multiple times and then edges between the same type of sockets are permuted, avoiding those permutations that create small girths.
To obtain a spatially coupled LDPC construction, $L$ smaller protographs of a standard $(l,r)$ LDPC code are first coupled
to create the SC-LDPC protograph following similar rules as in the  $(l,r,L)$ case:
each variable node at position $u$ is connected to a check node at positions $u, u+1,\ldots,u+l-1$. 
For instance, Fig. \ref{fig:protographcoupling} illustrates the construction of the coupled protograph that generates the $(3,6,L=3)$ ensemble.
Once the coupled protograph is created, a code sampled from the $(l,r,L)_{\P}$ ensemble with $M$ bits per position is generated by permuting  edges between $M/k$ copies of the coupled protograph, where $k$ is the number of bits per position in the coupled protograph.
E.g., we have $k=2$ in the $(3,6)$ protograph in Fig. \ref{fig:protographcoupling}.
The rate of the $(l,r,L)_{\P}$ ensemble can be computed from the protograph:
\begin{align}
\rate_{(l,r,L)_{\P}}&=1-\frac{k(L+l-1)}{L}
\end{align}
which also tends to the uncoupled rate  when $L\rightarrow\infty$.

%

\subsection{The $(l,r,L)_{\P}$ degree distribution}

\begin{figure}[htb]
\begin{center}
\renewcommand{\figurewidth}{0.6\columnwidth}
\renewcommand{\figureheight}{0.35\columnwidth}
\footnotesize
\includegraphics{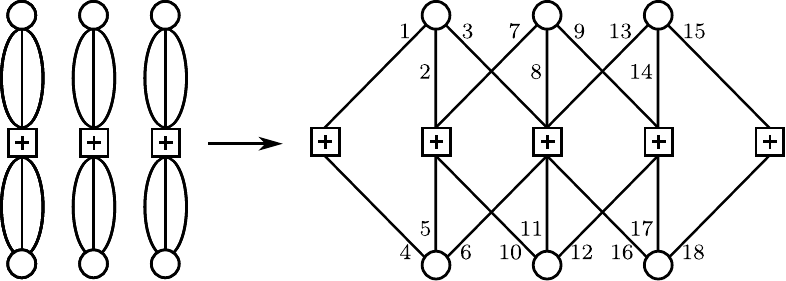}
\caption{Coupling of three $(3,6)$ protographs to create the \mbox{$(l,r,L)_{\P}=(3,6,3)_{\P}$} protograph.
The edges are numbered to refer to the multi-edge type notation.}
\label{fig:protographcoupling}
\end{center}
\end{figure}

In Section \ref{sec:peelingdecoder}, we compute the expected graph evolution under PD  protograph-based SC-LDPC ensembles.
As shown in \cite{Luby01}, the graph degree distribution (DD) at any time during the decoding process constitutes a sufficient statistic for analysis and thus we can reduce the problem to analyzing the average evolution of the degree distribution of the sequence of residual graphs.
The first step is to define a proper DD for the $(l,r,L)_{\P}$ ensemble, which is a particular class of multi-edge type LDPC codes.
Each edge of the coupled protograph constitutes a different edge type, e.g. as in Fig. \ref{fig:protographcoupling}, and thus the highest edge type index is $m=lkL$. 


We use a multi-edge like notation as introduced in \cite{Urbanke08-2}.
Denote a row vector with \vec{v} and a matrix with \M{M}.
\vec{1} denotes a vector consisting only of ones. $\vec{0}_{\sim i,j,k}$ is a vector of zeros except the entries $i,j,k$ which are $1$.
We denote with $|\vec{d}|=\sum_{i=1}^n d_i$ the sum of all entries of \vec{d}.
Suppose we generate a sample from the $(l,r,L)_{\P}$ ensemble.
For a given variable (check) node, we define its multi-edge type by a vector $\vec{d}=(d_1,\dots,d_m)$, where $d_j\in\{0, 1, \ldots\}$ represents the number of edges of type $j$ connected to the variable (check) node.
Let $\L_{\vec{d}}$ ($R_{\vec{d}}$) represent the number of variable (check) nodes of multi-edge type $\vec{d}$ and let $\Phi$ be the set of multi-edge types in the graph.
As in the case of single-type LDPC ensembles, the whole DD is represented in a compact way using multinomials
\begin{align}\label{eq:LR}
 \L(\vec{x})&=\sum_{\vec{d}\in\Phi} \L_{\vec{d}}\vec{x}^{\vec{d}},\qquad R(\vec{x})=\sum_{\vec{d}\in\Phi} R_{\vec{d}}\vec{x}^{\vec{d}}
\end{align}
where $\vec{x}^{\vec{d}}=\prod_{i=1}^{m}x^{d_i}_i$.
The total number of variable nodes and check nodes is given by $\L(\vec{1}), R(\vec{1})$.
For example, consider the  $(l,r,L)_{\P}=(3,6,3)$ ensemble in Fig. \ref{fig:protographcoupling}.
The multi-edge type distribution of the ensemble is given by
\begin{align}
 \L(\vec{x})&=\tfrac{M}{2}(x_1x_2x_3+x_4x_5x_6
+x_7x_8x_9 \nonumber
\\&+x_{10}x_{11}x_{12}\label{ex1}
+x_{13}x_{14}x_{15}+x_{16}x_{17}x_{18}),\\
R(\vec{x})&=\tfrac{M}{2}(x_1x_4+x_2x_5x_7x_{10}\nonumber\\
&+x_3x_6x_8x_{11}x_{13}x_{16}+x_9x_{12}x_{14}x_{17}+x_{15}x_{18}). \label{ex2}
\end{align}
Alternatively, we can specify the DD from an edge perspective:
\begin{align}\label{dev1}
\L_{x_j}(\vec{x})&=\frac{ d\L(\vec{x})}{dx_j} =\sum_{\vec{d}\in\Phi}   \L_{j,\vec{d}}\;  \vec{x}^{\left(\vec{d}-\vec{0}_{\sim j}\right)}\\\label{dev2}
R_{x_j}(\vec{x})&=\frac{ dR(\vec{x})}{dx_j}=\sum_{\vec{d}\in\Phi}  R_{j,\vec{d}}  \vec{x}^{\left(\vec{d}-\vec{0}_{\sim j}\right)}
\end{align}
where $\L_{j,\vec{d}}=d_j \L_{\vec{d}}$ ($ R_{j,\vec{d}} =d_jR_{\vec{d}}$) represents the number of edges of type $j$ for which the multi-edge type of the rest of the sockets of the variable (check) node is $\vec{d}$. 
Note that $\L_{x_j}(\vec{1})=R_{x_j}(\vec{1})$ gives the number of edges of a certain edge type. This notation is extended to allow multiple derivations, e.g. $R_{x_i,x_j,x_k}(\vec{x})$.
For $d_i,d_j,d_k\leq1$ we obtain the number of edges connected to nodes which have certain sockets.
%
%

%


\section{Expected Graph Evolution}\label{sec:peelingdecoder}
Consider transmission over the BEC and PD \cite{Luby01}.
We start the PD algorithm by removing the variable nodes and their edges associated with non-erased symbols and any disconnected check nodes from the graph; following this, one deg1 check node and its linked variable node are removed from the reduced graph per iteration.
In \cite{Luby01}, it was shown that  the sequence of graphs follows a typical path or expected evolution.
Based on the graph covariance evolution as a function of the code length at those points where the expected evolution of the fraction of deg1 check nodes presents a local minimum (critical points), \emph{scaling laws} (SLs)  predict  the finite-length performance of code ensembles in the waterfall region  in \cite{Urbanke09}.
In this section, we show how to compute the graph expected evolution for the $(l,r,L)_{\P}$ ensemble.

The PD is initialized by removing from the graph all bits correctly received.
Therefore, each variable node is removed from the graph with probability $(1-\epsilon)$.
Denote by $\L(\vec{x},t=0)$ and $R(\vec{x},t=0)$ the expected graph DD after this step.
It can be easily checked that
\begin{align}\label{L0}
\L(\vec{x},t=0)=\sum_{\vec{d}\in\Phi} \epsilon \L_{\vec{d}}\vec{x}^{\vec{d}}.
\end{align}

Given a check node of multi-edge type $\vec{d}$, define $\Set{D}(\vec{d})$ as the set of multi-edge types  which can be created from $\vec{d}$ after the transmission
\footnote{For instance, after transmission of the  $(3,6,3)_{\P}$ ensemble in \eqref{ex1} and \eqref{ex2}, check nodes of multi-edge type $\vec{d}'=\vec{0}_{1}$, $\vec{d}'=\vec{0}_{2}$ and $\vec{d}'=\vec{0}$ can be created from a check node of multi-edge type $\vec{d}=\vec{0}_{1,2}$.}.
It is straightforward to show that
\begin{align}\label{R0}
R(\vec{x},t=0)=\sum_{\vec{d}}R_{\vec{d}} \sum_{\vec{d}'\in \Set{D}(\vec{d})} \epsilon^{|\vec{d}'|}(1-\epsilon)^{|\vec{d}|-|\vec{d}'|} \vec{x}^{\vec{d}'}.
\end{align}


We define the extended set of all possible multi-edge types after transmission as $\overline{\Phi}=\Phi\bigcup_{\vec{d}}\Set{D}(\vec{d})$.
Let $\L_{\vec{d}}(\ell)$ and $R_{\vec{d}}(\ell)$ be the (random) DDs after $\ell$ iterations.
We define the normalized (random) DDs at normalized time $\tau$ as follows
\begin{align}\label{norm}
\tau\doteq\frac{\ell}{ M}, \qquad r_{\vec{d}}(\tau)\doteq\frac{R_{\vec{d}}(\ell)}{M}, \qquad \l_{\vec{d}}(\tau)\doteq\frac{\L_{\vec{d}}(\ell)}{M}
\end{align} 
and by extension
\begin{align}\label{rtau}
\l(\vec{x},\tau)&=\sum_{\vec{d}\in\overline{\Phi}} \l_{\vec{d}}(\tau) \vec{x}^{\vec{d}}, \qquad r(\vec{x},\tau)=\sum_{\vec{d}\in\overline{\Phi}} r_{\vec{d}}(\tau) \vec{x}^{\vec{d}}.
\end{align}
As shown in \cite{Luby01}, the mean values $\hat{\l}_{\vec{d}}(\tau)$ and $\hat{r}_{\vec{d}}(\tau)$ of this random process at time $\tau$ are given by the solution to the following system of differential equations:
\begin{align}
\frac{\partial \hat{\l}_{\vec{d}}(\tau)}{\partial \tau}&=\E[\L_{\vec{d}}(\ell+1)-\L_{\vec{d}}(\ell)\Big|\hat{v}(\vec{x},\ell), \hat{r}(\vec{x},\ell)]\label{system2}
\\
\frac{\partial \hat{r}_{\vec{d}}(\tau)}{\partial \tau}&=\E[R_{\vec{d}}(\ell+1)-R_{\vec{d}}(\ell)\Big|\hat{v}(\vec{x},\ell), \hat{r}(\vec{x},\ell)]\label{system1}.
\end{align}
Further, the solution is unique and, with probability \mbox{$1-\mathcal{O}(\text{e}^{-\sqrt{M}})$}, 
 any particular realization of the normalized DD in (\ref{norm}) deviates from its mean by a factor of less than $M^{-1/6}$ for the initial conditions $\hat{r}_{\vec{d}}(0)=\E[R_{\vec{d}}(\ell=0)]/ M$ and $ \hat{\l}_{\vec{d}}(0)=\E[\L_{\vec{d}}(\ell=0)]/M$, computed in \eqref{L0} and \eqref{R0}.
The expectations in \eqref{system1} and \eqref{system2} are given in Section \ref{subsec:meanev}. 
The ensemble BP threshold is given by the maximum $\pe$ for which the mean total fraction of deg1 check nodes, given by
\begin{align}\label{r1mean}
\hat{c}_1(\tau)\doteq\sum_{i=1}^m \hat{r}(\vec{0}_{\sim i},\tau)
\end{align}
is positive for any $\tau\in[0,\epsilon L]$, where $\hat{r}(\vec{x},\tau)$ is the mean of $r(\vec{x},\tau)$ in \eqref{rtau}.
Based on (\ref{r1mean}), we can say that $\hat{c}_{1}(\tau)$ is the mean of the random process $c_{1}(\tau)$.

\begin{figure}[th]
\renewcommand{\figurewidth}{0.7\columnwidth}
\renewcommand{\figureheight}{0.4\columnwidth}
\footnotesize
\begin{tabular}{c}
\includegraphics{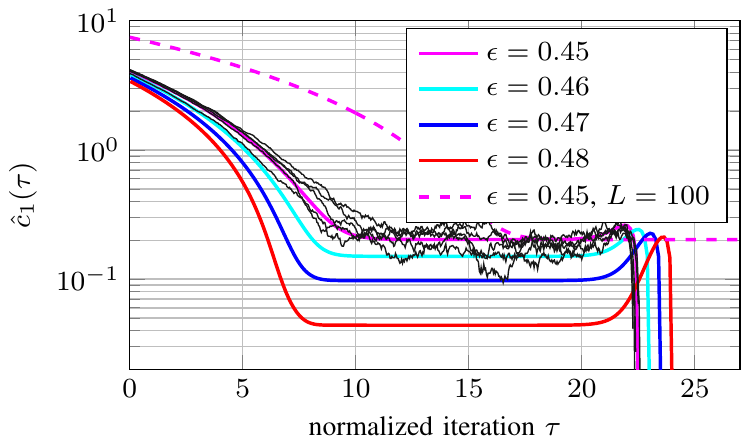} \\ (a) \\ 
\hspace*{0.8em}\includegraphics{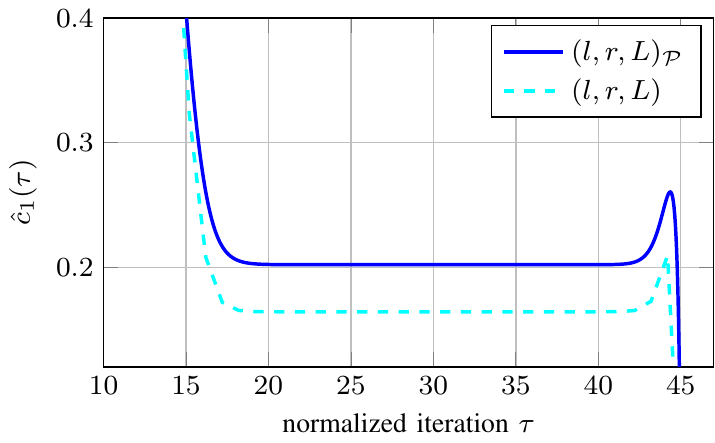}\\
(b)
\end{tabular}
\caption{ Subplot (a) shows $\hat{c}_1(\tau)$ for the  $(l,r,L)_{\P}=(3,6,50)_{\P}$ ensemble for a varying $\epsilon$.
For $\epsilon=0.45$, the subplot includes actual decoding trajectories. Subplot (b) shows $\hat{c}_1(\tau)$ for the ensembles $(l,r,L)_{\P}=(3,6,100)_{\P}$ (solid line) and $(l,r,L)$ (dashed line) and $\epsilon=0.45$.}\label{means}
\end{figure}

\subsection{Expected graph evolution in a single PD step}\label{subsec:meanev}
We need to calculate the expectations in \eqref{system1} and \eqref{system2}.
For the $(l,r,L)_{\P}$ ensemble,  there are no degrees 
higher than $1$, i.e. for all $\vec{d}\in\overline{\Phi}$, we have $d_j\leq 1$.
We start with the evolution of the number of check nodes of a given multi-edge type $\vec{d}\in\overline{\Phi}$.
At each iteration,  
 the probability that a  deg1 check node of edge type $j\in\{1,\ldots,m\}$ is directly removed is
\begin{align}
\D_{\vec{0}_{\sim j},\text{dir}}
=\frac{R(\vec{0}_{\sim j},\ell)}{\sum_{i=1}^m R(\vec{0}_{\sim i},\ell)}
=\frac{R(\vec{0}_{\sim j},\ell)}{C_1(\ell)}
\end{align}
where $C_1(\ell)$ is the total number of deg1 check nodes. Denote by $\D^{-}_{\vec{d},\text{indir}}$ the probability that a check node of multi-edge type $\vec{d}$ is lost because the node has lost one edge. 
We also obtain
\begin{align}
\D^{-}_{\vec{d},\text{indir}}=
\sum_{j=1}^m\D_{\vec{0}_{\sim j},\text{dir}}\hspace{-1mm}\sum_{k=1}^{m}\hspace{-1mm}
\frac{\L_{x_j,x_k}(\vec{1},\ell)}{\L_{x_j}(\vec{1},\ell)}
\frac{R_{\vec{d}}(\ell)}{R_{x_k}(\vec{1},\ell)}
\end{align}
where $\L_{x_j,x_k}(\vec{1},\ell)$ is the number of variable nodes that have sockets of type $j$ and $k$. Similarly, every time a check node of multi-edge type $\vec{d}$ is lost because we have removed one edge, e.g. an edge of type $j$, we create a new check node of multi-edge type $\vec{d}-\vec{0}_{\sim j}$.
Let  $\D^{+}_{\vec{d},\text{indir}}$ denote the probability that we create a check node of multi-edge type $\vec{d}$ after the PD iteration.
This probability is given by
\begin{align}
\D^{+}_{\vec{d},\text{indir}}
&=\sum_{k:d_k=0}
\D^{-}_{(\vec{d}+\vec{0}_{\sim k}),\text{indir}}.
\end{align}
Putting all together, we obtain
\begin{align*}
\E[R_{\vec{d}}(\ell+1)-R_{\vec{d}}(\ell)]&=\D^{+}_{\vec{d},\text{indir}}-\D^{+}_{\vec{d},\text{indir}}\\
\E[R_{\vec{0}_{\sim j}}(\ell+1)-R_{\vec{0}_{\sim j}}(\ell)]&=-\D_{\vec{0}_{\sim j},\text{dir}}+\D^{+}_{\vec{d},\text{indir}}-\D^{+}_{\vec{d},\text{indir}}
\end{align*}
for $j=1,\ldots,m$ and $|\vec{d}|>1$. Finally, \eqref{system2} is given by
\begin{align}
\E[\L_{\vec{d}}(\ell+1)-\L_{\vec{d}}(\ell)]=-\sum_{j: d_j=1}\frac{R(\vec{0}_{\sim j},\ell)}{C_1(\ell)}\frac{\L_{\vec{d}}(\ell)}{\L_{x_j}(\vec{1},\ell)}.
\end{align}


\subsection{Mean evolution of  degree-1 Check Nodes during PD}
The analytical calculation of $\hat{c}_1(\tau)$ for the protograph-based ensemble $(l,r,L)_{\P}=(3,6,50)_{\P}$ for a varying $\epsilon$ is shown in Fig. \ref{means}(a).
The BP threshold is $\epsilon_{(3,6,50)_{\P}}\approx 0.48815$.
As explained in detail in \cite{Olmos13}, in the initial phase deg1 check nodes are removed more or less uniformly along the chain.
In Fig. \ref{means}(a), this phase corresponds to the initial decreasing branch.
Now the second phase starts, which corresponds to the ``decoding'' wave that  moves at constant speed through the graph \cite{KudekarBMS}.
In this phase we do not have one critical time point at which the decoder is most likely to stop, but the expected number of deg1 check nodes is
essentially a constant.
Therefore, we call this phase the ``steady-state'' phase.
Denote by $\hat{c}_{1}(*)$ the value during such phase.
An important observation is that $\hat{c}_{1}(*)$ does not depend on the chain length $L$, as we can see in Fig. \ref{means}(a) where the curve for $L=100$ and $\epsilon=0.45$ is included.
As in \cite{Olmos13}, we have observed that $\hat{c}_{1}(*)$ is accurately estimated by a first-order Taylor expansion around the threshold $\epsilon^*$:
\begin{align}\label{gamma}
\hat{c}_{1}(*)\approx \gamma \Delta_{\epsilon}
\end{align}
where $\gamma$ is a constant that depends on the underlying regular $(l,r)$ ensemble and the way the coupled code is constructed.
For the $(3,6,L)_{\P}$ ensemble, we obtain $\gamma_{\P}\approx 5.25$.
In Fig. \ref{means}(b), we compare  the  $\hat{c}_{1}(\tau)$ evolution for  the  ensembles $(3,6,L=100)$ (dashed line) and $(3,6,L=100)_{\P}$ (solid line) for $\epsilon=0.45$.
While both ensembles have the same threshold $(\epsilon^*=0.48815)$,  $\hat{c}_{1}(*)$ is significantly lower for the random case,
which means the decoding process of the random ensemble is less robust against decoding failures.
Indeed, the $\gamma$ parameter in \eqref{gamma} for the $(3,6,L)$ ensemble is $\gamma\approx4.2<\gamma_{\P}$.


\section{Finite-length scaling behavior}\label{finite}
As shown in \cite{Olmos13,Olmos13-3} for the $(l,r,L)$ random ensemble, the process $c_1(\tau)$ during the steady-state phase converges in $M$ to a Gauss-Markov process with constant mean and variance and  exponentially decreasing time covariance, i.e. $\text{CoV}[c_1(\tau), c_1(\zeta)]\propto \exp(-\theta |\zeta-\tau|)$, where $\theta=\theta(l,r)$ depends on the coupling pattern \cite{Olmos13-3}.
Based on this behavior, the zero-crossing probability of the process $c_1(\tau)$ during the steady-state phase is estimated as follows:
\begin{align}\label{longchain}
P^*&\approx 1-\exp\left(-\frac{(\epsilon L-\tau^*)}{\mu_0(M,\epsilon,l,r)}\right)
\end{align}
where $(\epsilon L-\tau^*)$ is the duration of the steady-state phase and  $\mu_0$ is the average survival time of the $c_1(\tau)$ process:
\begin{align}\label{mu}
&\mu_0(l,r,M,\epsilon)\approx\frac{\sqrt{2\pi}}{\theta}\int_{0}^{ \frac{\sqrt{M}\Delta_\epsilon}{\alpha}}\Phi(z)\text{e}^{\frac{1}{2}z^2}\text{d}z
\end{align}
where  $\Phi(z)$ is the  c.d.f. of the  Gaussian $\mathcal{N}(0,1)$, $\alpha=\delta_1(\ast,\epsilon)\gamma^{-1}$ and $\delta_1(\ast)$ is proportional to the variance of $c_1(\tau)$. 
Note that while both $\theta$ and $\delta_1(\ast)$ have the same value for the $(l,r,L)$ and  $(l,r,L)_{\P}$ ensembles (we will see later that this assumption is accurate), the higher $\gamma$  obtained for the protograph case  yields an exponential increase of $\mu_0$ in \eqref{mu} and, consequently, a drastic reduction in the error performance estimate in \eqref{longchain}.

\begin{figure}[ht]
\renewcommand{\figurewidth}{0.7\columnwidth}
\renewcommand{\figureheight}{0.4\columnwidth}
\footnotesize
\begin{tabular}{c}
\hspace*{0.5em}  \includegraphics{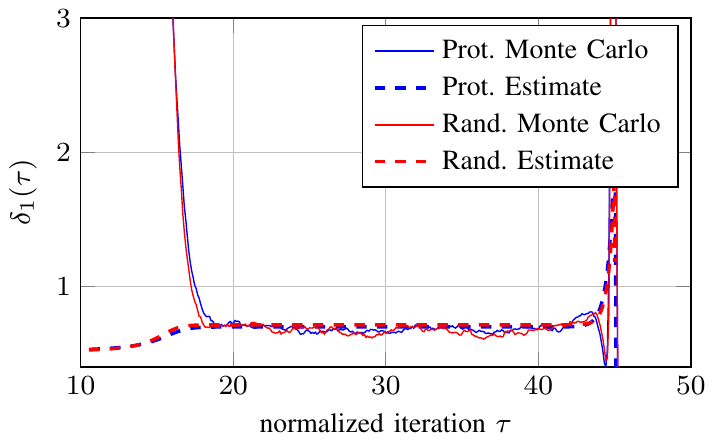}  \\ (a) \\ 
\includegraphics{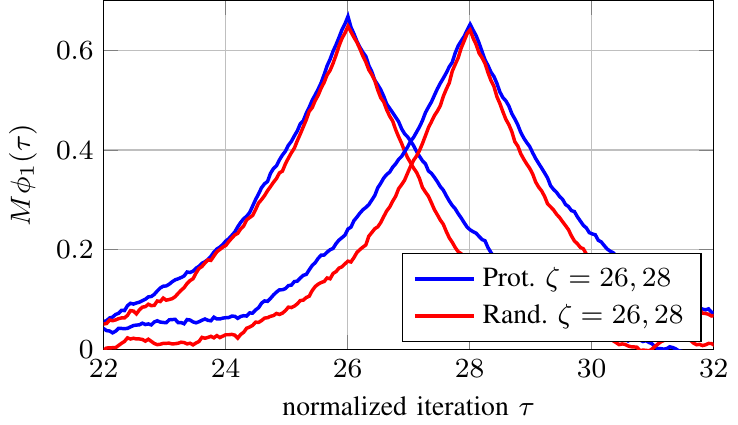}\\
(b)
\end{tabular}
\caption{Subplot (a) shows Monte Carlo and the proposed estimates to $\delta_1(\tau)$ for  the $(3,6,100)_{\P}$ and  $(3,6,100)$ ensembles with $M=2000$.
Subplot (b) shows the process covariance estimation at $2$ time instants  for the same ensembles.
All results are computed for $\epsilon=0.45$.}\label{vars}
\end{figure}

\subsection{Variance evolution}\label{var}

As shown in \cite{Urbanke09}, the PD covariance evolution, i.e., the time evolution of moments of the form
$\delta_{\vec{d},\vec{d}'}(\tau)\doteq\text{CoVar}[r_{\vec{d}}(\tau),r_{\vec{d}'}(\tau)]$ can be predicted by solving an extended system of differential equations. 
Further, $r_{\vec{d}}(\tau)$ $\forall \vec{d}\in\overline{\Phi}$ converges (in $M$) to a multivariate Gaussian distribution with mean $\hat{r}_{\vec{d}}(\tau)$ and a covariance matrix given by the  moments $\delta_{\vec{d},\vec{d}'}/M$.
In particular, we are interested in 
\begin{align}
\delta_1(\tau)=\sum_{i=1}^{m}\sum_{b=1}^{m}\delta_{\vec{0}_{\sim i},\vec{0}_{\sim b}}(\tau)
\end{align}
since $\text{Var}[c_{1}(\tau)]=\delta_1(\tau)/M$.
In Fig. \ref{vars}(a), we plot a Monte Carlo estimate to  $\delta_1(\tau)$  for the $(3,6,100)_{\P}$ and  $(3,6,100)$ ensembles with $M=2000$. For both ensembles, $\text{Var}[c_{1}(\tau)]$ in the steady-state phase remains approximately constant
and we denote this constant value by $\delta_1(\ast)$.
In addition, there is no significant mismatch observed between $\delta_1(\ast)$ and $\delta_1(\ast)_{\P}$. 


\subsection{A simple estimate of $\delta_1(\ast)$}\label{est}
Instead of solving the covariance evolution for the $(l,r,L)_{\P}$ ensemble analytically, which requires to numerically integrate a system of $\mathcal{O}(L^2)$ differential equations \cite{Olmos13-3}, we present a method  to accurately estimate $\delta_1(\ast)_{\P}$ using only the expected graph evolution.
This method can also be applied to the random $(l,r,L)$ ensemble.
Assume the  graph has followed the exact mean evolution path up to an iteration $\ell-1$.
In the next iteration, the graph evolution is assumed to be random. 
Let $\Delta_1(\ell)=C_1(\ell)-\widehat{C}_1(\ell-1)$ be the random evolution in the number of deg1 check nodes after we perform the next PD iteration,
where by assumption only $C_1(\ell)$ is a random variable.
Note that $\Delta_1(\ell)\in\{-l,\dots,l-1\}$.  We propose to approximate $\delta_1(\tau=\ell/M)$ with the variance of $\Delta_1(\ell)$, i.e., 
\begin{align}\label{eq:deltaestimate}
\delta_1(\tau)\approx\text{Var}[\Delta_1(\tau)]=\left(\E[\Delta_1(\tau)^2]-\E[\Delta_1(\tau)]^2\right).
\end{align}
Computing the probability of all possible outcomes of $\Delta_1(\tau)$ is done by similar techniques to those applied in Section \ref{subsec:meanev}.
In Fig. \ref{vars} (a), the variance estimate in \eqref{eq:deltaestimate} is plotted along with the Monte Carlo estimate to $\delta_1(\tau)$.
As observed, we achieve an accurate estimate during the steady-state phase that can be efficiently computed for both ensembles.
We are still exploring the reasons why our proposal provides such a good estimate in the case of SC-LDPC codes, in contrast to the uncoupled LDPC case.
We conjecture that the fact that $\hat{c}_1(\tau)$ has a constant evolution during the steady-state phase plays a major role to explain our result.
The curvature of the mean evolution curve, which is zero in our case, at the critical points has been related to the variance parameter during the analysis of finite-length uncoupled LDPC codes \cite{Ezri08}.

\subsection{Process covariance at two time instants}\label{covdecay}
As errors happen more or less uniformly through the  steady-state phase, 
we finally need to estimate the process covariance with time, i.e.
$\phi_{1}(\tau,\zeta)\doteq\E[c_1(\tau)c_1(\zeta)]-\hat{c}_{1}(\tau)\hat{c}_1(\zeta)$.
We take an empirical approach to determine this quantity.
In Fig. \ref{vars}(b), we plot the Monte Carlo estimate of $M \phi_{1}(\tau,\zeta)$ for
$\zeta=26$ and $\zeta=28$ for the random and the protograph ensembles using $2000$ decoding trajectories for $\epsilon=0.45$ and $M=2000$.
As observed, during the steady-state period, the covariance decay is exponential with $|\zeta-\tau|$:
\begin{align}
\phi_{1}(\tau,\zeta)\approx \frac{\delta_1(\ast)}{M} \text{e}^{-\theta
|\zeta-\tau|}
\end{align} 
and we get the same decay rate $\theta\approx0.6$ for both cases.

\section{Performance comparison}
\begin{figure}[htb]
\renewcommand{\figurewidth}{0.8\columnwidth}
\renewcommand{\figureheight}{0.7\columnwidth}
\footnotesize
\includegraphics{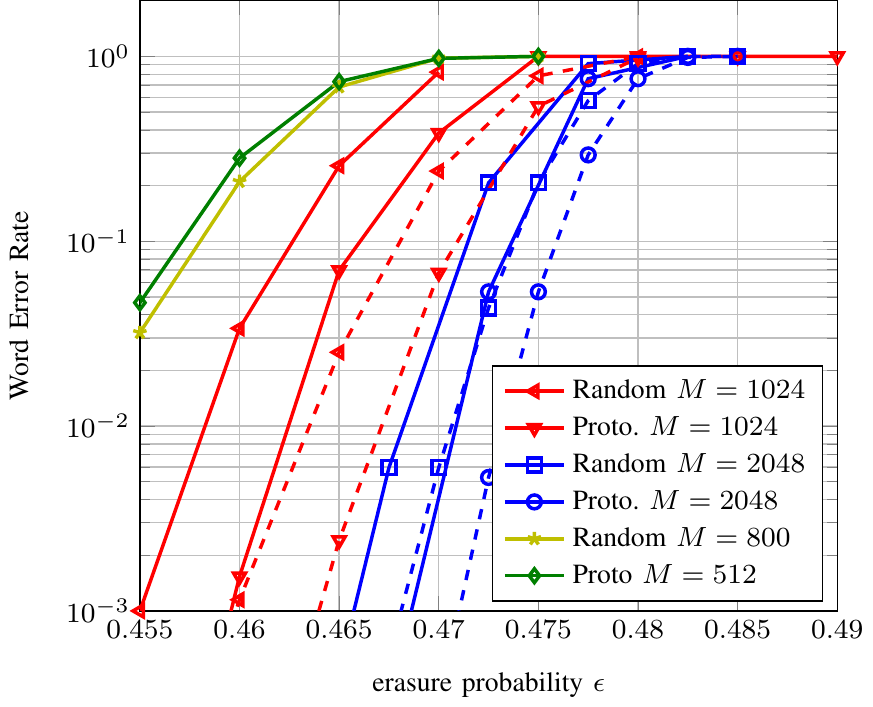}
\caption{Word error rate on the BEC for the $(l,r,L)=(3,6,100)$ and \mbox{$(l,r,L)_{\P}=(3,6,100)_{\P}$} ensembles.}
\label{WERs}
\end{figure}

In Fig. \ref{WERs}, we show performance results for the $(3,6,100)$ and $(3,6,100)_{\P}$ ensembles.
Solid lines represent actual error rates computed by Monte Carlo simulations while dashed lines correspond to the performance estimate given by the scaling law in \eqref{longchain} and \eqref{mu}.
For the same rate and code length, the protograph-based codes have error rates one order of magnitude below the random case.

As discussed in \cite{Olmos13,Olmos13-3}, the scaling law proposed for SC-LDPC codes has a decreasing (in $M$) shift with respect to the actual performance.
Nonetheless, it captures the right scaling between code performance, code parameters and the gap to the threshold.
In light of the results in  Fig. \ref{vars}(a) and Fig. \ref{vars}(b), the performance gain obtained  is essentially explained by the height of the mean evolution $\hat{c}_{1}(\tau)$ during the steady-state phase.
This performance gain can be simply predicted by evaluating the effect of $\gamma$ in \eqref{longchain} and \eqref{mu}.
Conversely, the increase in the number $M$ of bits per position that the $(l,r,L)$ ensemble would require to achieve the $(l,r,L)_{\P}$ performance can be computed using \eqref{mu}.
For instance, for the $(3,6,L)$ ensemble, we have $\gamma_{\P}/\gamma\approx 1.25$, which by \eqref{mu} implies that we have to multiply 
$M$ by $1.57$  in the random  ensemble to match the $(3,6,L)_{\P}$ performance.
In Fig. \ref{WERs}, we include the simulation performance curve for the $(3,6,100)$ ensemble with $M=800\approx 1.57\cdot512$, as we show it essentially fits the $(3,6,100)$ performance with $M=512$.

\section{Conclusion \& Outlook}
We have shown that  protograph ensembles significantly improve performance in the waterfall region.
The scaling law for protograph-ensembles can be used to help the code design process.
Using these tools, we plan to 
compare SC-LDPC ensembles based on the same $(l,r)$-regular LDPC code using different protograph-based matrices.
We also suggest to analyze why the covariance estimate works so well.

%
%
%
%



%

%

\bibliographystyle{IEEEtran}
\bibliography{IEEEabrv,confs-jrnls,allbib20_03.bib}

\end{document}